\documentclass[conference,10pt,twocolumn]{IEEEtran}

\usepackage{amsmath}
\usepackage{amsthm}
\usepackage{amssymb}
\usepackage{graphicx}
\usepackage{subfig}
\usepackage{rotating}
\usepackage{flushend}
\usepackage{verbatim}
\allowdisplaybreaks

\newsavebox{\tempbox}

\usepackage{xcolor}

\newcommand{\bad}{\boldsymbol{!}}

\newcommand{\logt}{\log_{2}}

\newcommand{\abs}[1]{\left| #1 \right|}

\newcommand{\typical}[1]{T^{n}_{\left[ #1 \right] }}
\newcommand{\type}[1]{T^{n}_{ #1  }}

\newcommand{\mcf}[1]{\mathcal{#1}}
\newcommand{\idc}[1]{\mathfrak{1}\left({#1}\right) }

\newcommand{\set}[1]{\left\{ #1 \right\}}

\DeclareMathOperator{\markov}{\setlength{\unitlength}{.5cm} \begin{picture}(1,1)  \put(0,.22){\line(1,0){1}}  \put(.5,.22){\circle{.3}}   \end{picture}}

\newtheorem{theorem}{Theorem}
\newtheorem{define}[theorem]{Definition}

\newtheorem{remark}[theorem]{Remark}
\let\oldremark\remark
\renewcommand{\remark}{\oldremark\normalfont}

\IEEEoverridecommandlockouts

\begin{document}

\title{Keyless authentication in the presence of a simultaneously transmitting adversary}

\author{\IEEEauthorblockN{Eric Graves} \\
\IEEEauthorblockA{Army Research Lab \\
Adelphi, MD 20783, U.S.A. \\
\texttt{ericsgra@ufl.edu}}
\and
\IEEEauthorblockN{Paul Yu} \\
\IEEEauthorblockA{Army Research Lab \\
Adelphi, MD 20783, U.S.A. \\
\texttt{paul.l.yu.civ@mail.mil}}
\and
\IEEEauthorblockN{Predrag Spasojevic} \\
\IEEEauthorblockA{WINLAB, ECE Department\\
 Rutgers University\\
\texttt{spasojev@winlab.rutgers.edu}}
}

\maketitle

\begin{abstract}
If Alice must communicate with Bob over a channel shared with the adversarial Eve, then Bob must be able to validate the authenticity of the message. In particular we consider the model where Alice and Eve share a discrete memoryless multiple access channel with Bob, thus allowing simultaneous transmissions from Alice and Eve. By traditional random coding arguments, we demonstrate an inner bound on the rate at which Alice may transmit, while still granting Bob the ability to authenticate. Furthermore this is accomplished in spite of Alice and Bob lacking a pre-shared key, as well as allowing Eve prior knowledge of both the codebook Alice and Bob share and the messages Alice transmits.
\end{abstract}

\section{Introduction}

In this paper we study physical layer authentication over a noisy multiple access channel and no pre-shared key. Such scenarios may occur when a network is established in a hostile environment. Being able to trust the information observed is just as important as actually observing it. It also stands to reason, that always having a pre-shared key seems restrictive as it is dependent on a method to generate and secure the key. We model this scenario by considering two transmitters (Alice and ne'er-do-well Eve) and one receiver (Bob). Alice and Eve share a discrete memoryless multiple access channel (DM-MAC), so that not only may Eve try to forge messages to Bob, she may try to modify the ones Alice transmits. To further complicate the matter, we assume that Eve not only knows the codebook which Alice and Bob share, but the message Alice will send as well. Our goal is for Bob to be able to validate the authenticity of the sender (or alternatively the integrity of the information), as well as be able to decode the information given it is authentic.

Simmons first studied physical layer authentication in~\cite{Auth}. Simmons allowed for Alice and Bob to have a pre-shared key, which Eve did not have access to. This key, which can be made small relative to the number of symbols it protects, in essence partitions the entire state of messages into sets which are valid and ones that are not. Any attempt by Eve to modify the message then results, with high probability, in a message included in the invalid set.

Physical layer authentication, without the use of pre-shared key, was considered by Maurer in~\cite{Binning}. Instead of a pre-shared key, Alice, Bob, and Eve are given given access to different random variables, which share some arbitrary correlation. By using their individual random variables, Alice and Bob proceed in public discussion over a perfect channel to generate a secret key. This is related to authentication in that the channel being perfect allows Eve the ability to perfectly spoof any packet from Alice. Hence, the distributions of the random variables play a pivotal role in establishing the authenticity of any message. In this model the three nodes must share copies of a random variable generated from an independent and identically distributed (iid) source. And furthermore the variables that Alice and Bob share must not be ``simulatable'' by Eve given her random variable. Use of an iid source though guarantees no underlying code structure, which could leak information about the random variables Alice and Bob view. It is also impractical in scenarios not concerned with secrecy to first generate a secret key for authentication purposes in later transmissions.

Because of these limits others have studied authentication in noisy channels. Works include Yu et al. who in~\cite{Yu08} used spread spectrum techniques in addition to a covert channel to ensure authentication~\cite{Yu08}. And Xiao et al in~\cite{Xiao08}, who used the unique scattering of individual users in indoor environments to authenticate packets. 
%As well as Korzhik et al in~\cite{Korzhik07}, who makes use of an initialization setup which may be noisy to create unique correlations which can be used for authentication. 
For purpose of discussion we focus on two works in particular. First, Lai et al., who in~\cite{Lai09} considered the problem of authentication in a noisy channel model when Alice and Bob have a pre-shared key. In their model the link between Bob and Alice is noisy while the link between Eve and Bob is noise-free, a worst case scenario. Much like the work of Simmons, the key Alice and Bob share is used to label some transmitted sequences as valid and not. And much like Simmons work, it is then nearly impossible for Eve to pick a transmitted message that would be valid. The second work, is that of Jiang~\cite{jiang2014keyless}. Jiang in specific considered the case where either Alice or Eve may transmit to Bob, but not simultaneously. By assuming the channels distinct, Jiang showed that even without a pre-shared secret key it was possible to guarantee authentication. This was done by considering the intersection of the distributions Alice could induce at Bob and the distributions Eve could induce at Bob had only marginal overlap if the channel from Alice to Bob was not simulatable using the channel from Eve to Bob. 

In contrast, by allowing simultaneous communication, we allow more possibilities for Eve's attack. This is because Eve does not need to attack the entire message transmitted, only a smaller portion, and thus only slightly disturb the distributions, which as mentioned previously were critical to ensuring authentication. Much like Jiang, we do this by showing that the distributions induced by the legitimate party, and the ones by the illegitimate party have marginal overlap. Unlike Jiang though, we approach the problem using traditional random coding methods and techniques. We derive a new inner bound, which is dependent on the channel not having a simulatable-like property. Unlike simulability though, this property can be controlled in part by Alice and her choice of distribution.

%ntegrity~\cite{graves2013coding,gravesDiss}. 

\section{Notation}\label{sec:notation}

Random variables will be denoted by upper case letters, their alphabets will be the script version of that letter, and particular realizations of that random variable will be denoted in lower case. In other words, $X$ is a random variable, and $x \in \mcf{X}$ is a particular realization of $X$. The $n$-th extension of a random variable is denoted $X^n$, and particular realizations of that random variable are denoted similarly by $x^n$. To represent a particular value in the $n$-th extension we use subscripts, such as $x_i$ is the value in the $i$-th position of $x^n$. 

When the need arises for discussion of the actual distribution of the random variable over their particular alphabet, we will use $P$ with that related random variable as a subscript ($P_X$ is the distribution $X$ has over $\mcf{X}$). When the distribution is clear from context, we drop the subscript. By $X\markov Y \markov Z$ we denote when $X,Y,Z$ form a Markov chain $\left(P_{Z,Y|X}(z,y|x) = P_{Z|Y}(z|y) P_{Y|X}(y|x)\right)$. 

By $\typical{X}$ we define the set $\left\{ x^n : \sum_{x \in \mcf{X}} \abs{P_{x^n}(x) - P_{X}(x) } \leq \delta_n \right\}$, where $P_{x^n}(x)$ is the induced type of $x^n$. Generally, $\typical{X}$ is referred to as a strongly typical set. We assume that $\delta$ follows the delta convention outlined in~\cite[Convention~2.11]{CK}. Furthermore by $H(X)$ and $H(X|Y)$ we define the entropy (conditional entropy resp.) of random variable $X$ ($X$ given $Y$ resp.), where $H(X) = -\sum_{x} P(x) \logt P(x)$ and $H(X|Y) = -\sum_{x,y} P(x,y) \logt P(x|y)$. Finally by $I(X;Y)$ we denote the mutual information of random variables $X$ and $Y$, where $I(X;Y) = \sum_{x,y} P(x,y) \logt \frac{P(x,y)}{P(x)P(y)}$.

\section{Channel Model}\label{sec:chan}

\begin{figure}
\centering 
\includegraphics[width=0.48\textwidth]{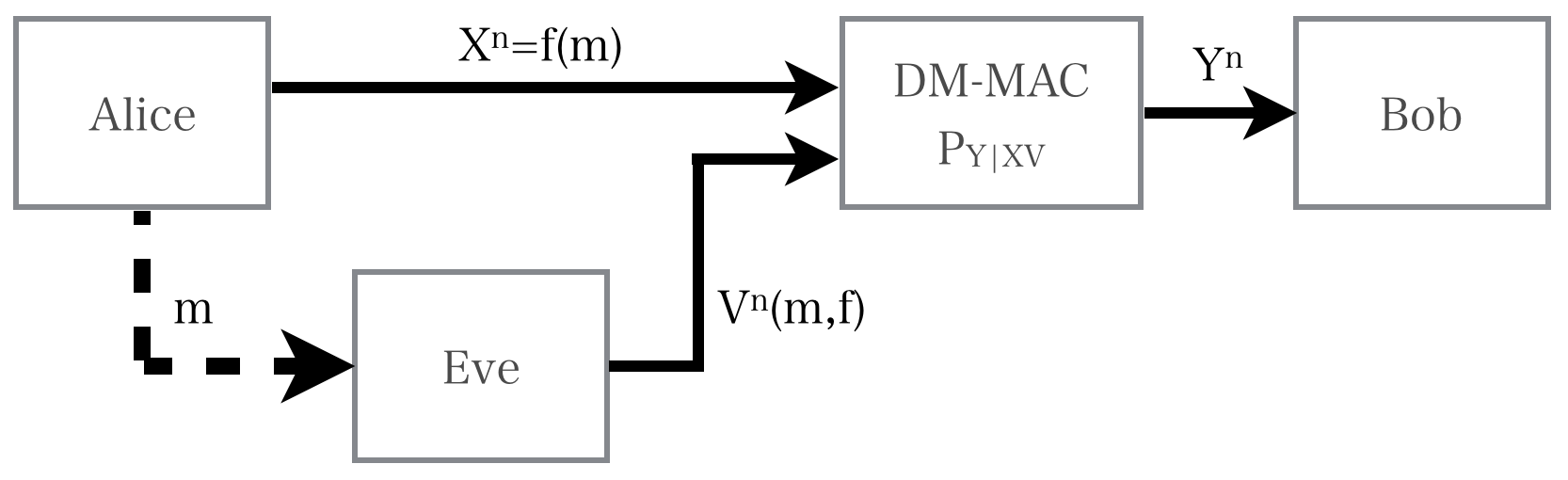}
  \caption{Channel model}
  \label{fig:v1}
\end{figure}

For our channel model, graphically represented in Figure~\ref{fig:v1}, Alice and Eve share a DM-MAC $(\mcf{X}\times \mcf{V}, P_{Y|X,V}, \mcf{Y} )$ with Bob. Alice wishes to transmit a message $m \in \mcf{M}$, while Eve attempts to modify its content so that Bob will believe Alice transmitted some $\hat m \in \mcf{M} $ that is not equal to $m$. By $n$ we denote the number of symbols that will be transmitted over the channel. Alice uses a possibly stochastic function $f: \mcf{M} \rightarrow \mcf{X}^n$ to generate her sequence which she transmits. Eve simultaneously chooses a sequence $v^n \in \mcf{V}^n$ as a function of the message and encoder, which she transmits concurrently with Alice's transmission. We assume that there exists an element $\emptyset \in \mcf{V}$ representative of the state when Eve decides not to transmit (hence Eve not attacking is denoted $V^n = \emptyset^n$.) Because the distribution associated with the output of the MAC when Eve is silent is of distinct importance, we refer to it specifically by $P_{Y|X,\emptyset}(x)$. In general, Bob will receive a value $y^n \in \mcf{Y}^n$, and then will use a decoder $\varphi: \mcf{Y}^n \rightarrow \mcf{M} \cup \set{\bad}$ to either output an estimate of the message, or instead to declare that Eve is attempting to intrude. We assume that Alice, Bob and Eve have access to the encoder $f$ and the decoder $\varphi$.

\section{Theorems and Definitions}\label{sec:def}

We expect the code to decode to the correct message when Eve is not attacking ($V^n = \emptyset^n$), and to either decode to the correct message or detect manipulation when Eve is attacking ($V^n \neq \emptyset^n$). These expectations lead directly lead to the following definition for our physical layer authentication code.
\begin{define}
A $(n,R,\epsilon_1,\epsilon_2)$-physical layer authentication code is any encoder-decoder pair $(f, \varphi)$, for which the encoder takes $2^{nR}$ messages and the decoder satisfies 
\begin{align*}
 \Pr \left\{ \varphi(Y^n) \neq M \middle| V^n = \emptyset^n \right\}  &\leq \epsilon_1 \\
 \Pr \left\{ \varphi(Y^n) \neq M \cup \bad \middle|  V^n \neq \emptyset^n \right\}  &\leq \epsilon_2.
\end{align*}
\end{define}
A code which meets these two requirements, can both decode and authenticate with high probability. We do not concern ourselves with the optimization of these two probabilities, and instead focus first on guaranteeing existence. As we will show using random coding, there exist codes such that both probabilities go to zero as $n$ grows to infinity. Section~\ref{sec:achiev} establishes the following inner bound on the rate of a physical layer authentication code.
\begin{theorem}
\[
\lim_{ (n,\epsilon_1,\epsilon_2)  \rightarrow (\infty,0,0)   } \max_{R: \exists \text{ a } (n,R,\epsilon_1,\epsilon_2) \text{ code } } R \geq \max_{U \in \mcf{U}^+ } I(Y;U|V = \emptyset)
\]
where $\mcf{U}^+$ contains all $U$ such that 
\begin{itemize}
\item $P_{Y,X,U}(y,x,u) = P_{Y|X,\emptyset}(y|x) P_{X|U}(x|u) P_{U}(u)$
%\item $P_{Y,U,V,X}(y,u,v,x) =  $
%\item $P_{Y|U,\emptyset} = P_{Y|X, \emptyset} P_{X|U}$
\item $ \sum_{x,\tilde u,v} P_{Y|X,V}(y|x,v) P_{X|U}(x|u) P_{U,V|U'}( u,v|u') \neq \sum_{x} P_{Y|X,\emptyset}(y|x) P_{X|U}(x|u')$ for all distributions $P_{U,V,U'}$ such that $P_{U}(u) = P_{U'}(u)$.
\end{itemize}
\end{theorem}
If the distribution $P_{X}$ which achieves capacity is a member of $\mcf{U}^+$, then the inner bound is also an outer bound. In general though, this is not the case. The necessity of auxiliary random variable $U$ can be demonstrated through a simple example. Let $\mcf{Y} = \set{ 0,1,2}$, $\mcf{X} = \set{0,1}$ and $\mcf{V} = \set{ \emptyset, 0,1}$, and define the DM-MAC by 
\[
P_{Y|X,V} = \left( \begin{matrix}. 9  & .1  & .1 & 0 & 0 & .9\\ .1 &.9 &.9 & 0 & 0 & .1  \\0 & 0 & 0 & 1 & 1 & 0  \end{matrix} \right)  
\]
where the realizations of $X,V$ are enumerated $(0,\emptyset),~ (1, \emptyset),~(0,0),~(1,0),~(0,1),~(1,1)$. For this channel though, choosing
\[
P_{X,V|\hat X} = \left( \begin{matrix} P_{X|\hat X}(0|0) & 0 \\ 0 & P_{X|\hat X}(1|1) \\ 0 & P_{X|\hat X}(0|1) \\ 0 & 0 \\ 0 & 0 \\ P_{X|\hat X}(1|0) & 0 \end{matrix} \right),
\]
results in 
\[
P_{Y|X,V} P_{X,V| \hat X} =  \left( \begin{matrix} .9 & .1 \\ .1 & .9 \\ 0 & 0\end{matrix} \right)  = P_{Y| X, \emptyset}.
\]
Thus $P_{X} \notin \mcf{U}^+$ for all distributions $P_{X}$. Now, instead lets introduce the auxiliary random variable $U$, and set $\mcf{U} = \set{0,1}$, with $P_{X|U}(0|0) = P_{X|U}(1|1) = p$ and $P_{X|U}(0|1) = P_{X|U}(1|0) = 1-p$. It is clear that if $Y=2$ then Bob can detect Eve, which corresponds to the case $\sum_{x, u,v} P_{Y|X,V}(2|x,v) P_{X|U}(x|u) P_{U,V|U'}(u,v|u') > 0$. In specific the value $2$ results with probability $1$ if $(x,v)$ equal $(0,1)$ or $(1,0)$ thus Eve must have $\sum_{u} P_{X|U}(1|u) P_{U,V|U'}(u,0|u') = \sum_{u} P_{X|U}(0|u) P_{U,V|U'}(u,1|u') = 0$. Clearly, assuming $p < .5$, $P_{X|U}(1|u) \geq p$ and $P_{X|U}(0|u)\geq p$. Thus it follows that $P_{V|U'}(0|u') = P_{V|U'}(1|u') = 0$. By adding an auxiliary random variable, we have now made the inner bound non-empty and thus increased our inner bound region. In this example it is not hard to see that in fact we can let $p \rightarrow 0$ and approach capacity.

Finally, in this work we assume that Eve knows the message $m$ and encoding function $f$. Why not allow for Eve to know $x^n$ as well? While this does in fact make Eve more powerful, it actually simplifies the math. Indeed allowing Eve knowledge of $x^n$ results in the property that for every $(n,R,\epsilon_1,\epsilon_2)$-physical layer authentication code there exist a corresponding deterministic $(n,R,\epsilon_1',\epsilon_2')$ physical layer authentication code such that $\epsilon'_1 + \epsilon'_2 \leq \epsilon_1 + \epsilon_2$. This follows because for any $M$
\begin{align*}
&\Pr \left\{ \varphi(Y^n) \neq M \middle| V^n = \emptyset^n, M=m \right\}  +  \\
&\hspace{35pt} + \Pr \left\{ \varphi(Y^n) \neq M \cup \bad \middle|  V^n \neq \emptyset^n \right\} \\
&= \sum_{x^n} P(x^n| m) \left( \Pr \left\{ \varphi(Y^n) \neq M \middle| V^n = \emptyset^n,X^n = x^n \right\}  \right. \\
&\hspace{35pt} \left.  +  \Pr \left\{ \varphi(Y^n) \neq M \cup \bad \middle|  V^n \neq \emptyset^n, X^n = x^n \right\} \right) \\
&\geq \min_{x^n: P(x^n|m) > 0 } \left( \Pr \left\{ \varphi(Y^n) \neq M \middle| V^n = \emptyset^n,X^n = x^n \right\}  \right. \\
&\hspace{35pt} \left.  +  \Pr \left\{ \varphi(Y^n) \neq M \cup \bad \middle|  V^n \neq \emptyset^n, X^n = x^n \right\} \right) .
\end{align*}
Thus the existence of a sequence of rate $R$ codes such that $\epsilon_1$ and $\epsilon_2$ converge to $0$, implies the existence of a sequence of rate $R$ deterministic codes for which $\epsilon_1'$ and $\epsilon_2'$ converge to $0$ as well. Hence considering deterministic encoders is sufficient for determining the maximum rate. Restricting the codes in Section~\ref{sec:achiev} results in the same conditions except without the auxiliary $U$, with $X$ instead replacing it. Our inner bound then is strictly smaller, as one would expect, due to the arguments of our example.

\section{Achievability}\label{sec:achiev}

\subsection{Construction}

Fix a distribution $P_{U,X}$. Define $R = \frac{1}{n} \logt{\abs{\mcf{M}}}$, and for all $m \in \mcf{M}$ randomly and independently select  $u^n(m) \in \type{U}$. For transmission, Alice chooses a value of $x^n$ according to the distribution $P(x^n | u^n(m) ) = \prod_{i=1}^{n} P(x_i | u_i(m) )$, and sends the value over the DM-MAC, $P_{Y|X,V}$. Bob, who receives a sequence $y^n$ from the channel, uses the decoder defined by 
\begin{align*}
\varphi(y^n) = \begin{cases} 
m &\text{ if } y^n \in \typical{Y|U,\emptyset}(u^n(m))  \\
&\text{and } y^n \notin   \bigcup_{ \hat m (\neq m) \in \mcf{M} } \typical{Y|U,\emptyset}(u^n(\hat m)) \\
\bad &\text{ otherwise},
\end{cases}
\end{align*} 
to estimate the message or declare intrusion.

For the remainder of the paper, by $U^n$ we will denote the random variable over the codebooks. Any particular $u^n$ is itself a function which defines a distribution over $x^n$ as given earlier.

\subsection{Reliability under no attack}

Given that $V^n = \emptyset^n$, an error will occur if either $y^n \notin \typical{Y|U,\emptyset}(u^n(m))$ or if $y^n \in \bigcup_{ \hat m (\neq m) \in \mcf{M} } \typical{Y|U,\emptyset}(u^n(\hat m))$. By using the union bound we may individually consider the probabilities of the two events. For the first event there exists an $\epsilon_n \rightarrow 0$ such that
\begin{equation}
\Pr \left\{ Y^n \notin \typical{Y|U,\emptyset}(u^n(m)) \middle| V^n = \emptyset^n  \right\} \leq \epsilon_n ,
\end{equation}
by~\cite[Lemma~2.12]{CK}. For the second, the existence of an $\epsilon_n' \rightarrow 0$ for which 
\begin{align}
\Pr \left\{ Y^n \in \bigcup_{ \hat m (\neq M) \in \mcf{M} } \typical{Y|U,\emptyset}(u^n(\hat m)) \right\} \leq \epsilon_n'
\end{align}
whenever $ R< I(Y;U) - \lambda_n $, for some $\lambda_n \rightarrow 0$, can be shown through traditional random coding arguments. 

\subsection{Detection}

%The major difference between our scheme and a code designed for transmission without authentication is that all detectable errors are labeled as malicious. Thus we must determine, given a particular $u^n$, if indeed Eve can choose a value of $v^n$ for which $\typical{Y|U,V}(u^n(m),v^n) \subset \typical{Y|U,V}(u^n(\hat m), \emptyset^n)$ for some $\hat m \neq m$. Clearly this property is based upon the DM-MAC being used for transmission. Suppose that $\typical{Y|U,V}(u^n(m),v^n) \cap \bigcup_{\hat m \neq m} \typical{Y|U,V}(u^n(\hat m), \emptyset^n)$ is small when compared to $\typical{Y|U,V}(u^n(m),v^n)$, it is then likely that that the received value of $y^n$ will indeed not like in any set and be detected. Indeed our code construction relies on this fact to guarantee authentication.

For any codebook, the value of $\Pr \left\{ \varphi(Y^n) \neq M \cup \bad \middle|  V^n \neq \emptyset^n \right\}$ is fixed. We want to show that the average over all codebooks results in $ \Pr \left\{ \varphi(Y^n) \neq M \cup \bad \middle|  V^n \neq \emptyset^n \right\} \rightarrow 0$ as $n$ increases. If the average probability of error for both the cases when Eve attacks and when she does not, converge to $0$, then there must exist a sequence of codes where the individual errors converge to $0$ as well.

For the probability of error given Eve attacking
\begin{align}
&E_{U^n}\Pr \left\{ \varphi(Y^n) \neq M \cup \bad \middle|  V^n \neq \emptyset^n \right\} \notag \\
&\leq \Pr \left\{ Y^n \in \bigcup_{\hat m \neq M } \typical{Y|U,\emptyset}( U^n(\hat m)) \middle| V^n \neq \emptyset^n  \right\}\notag \\
&= E_M \!\! \Pr \!\!\left\{ \!\! Y^n \!\!\in \!\! \bigcup_{\hat m \neq m } \typical{Y|U,\emptyset}( U^n(\hat m)) \middle| V^n \neq \emptyset^n, M=m  \right\}.
\end{align}
By symmetry we need only consider a particular value of $m \in M$, and hence we focus on
\begin{align}
&\Pr \left\{ Y^n \in \bigcup_{\hat m \neq m } \typical{Y|U,\emptyset}( U^n(\hat m)) \middle| M=m, V^n \neq \emptyset^n  \right\}. \label{eq:d_1}
\end{align}
The probability of error in this case is dependent on what Eve decides to transmit, which as mentioned previously is dependent upon the codebook, $u^n$, and the actual message transmitted $m$. For every codebook then, it is important to consider the probability of a value particular $y^n$ occurring given the codebook, message, and sequence transmitted. This is made explicit by writing equation~\eqref{eq:d_1} as
\begin{align}
&\sum_{u^n,v^n,y^n } \!\!\!\!\!\! P(y^n,v^n,u^n|M=m) \idc{y^n \!\! \in \!\!\!  \bigcup_{\hat m \neq m  } \!\! \typical{Y|U,\emptyset}( u^n(\hat m))} \label{eq:d_2}.
\end{align}
%It is important to note, that showing the convergence of equation~\eqref{eq:d_1}, is equivalent to showing that 
%\[
%E_{U^n} \Pr \left\{ Y^n \in \bigcup_{\hat m \neq m  } \typical{Y|U,\emptyset}(u^n(\hat m)) \middle| M=m  \right\} \rightarrow 0,
%\]
%for recall that $U^n$ is a random variable distributed over all possible codebooks. The probability of intrusion is clearly a function of our codebook, and the actions taken by Eve and Alice. But, in equation~\eqref{eq:d_1}, is the added element of $y^n$. In general it is possible that all possible attacks have a vanishing but non zero probability of intrusion. We would thus like to characterize the probability of intrusion based entirely on the codebook, the transmitted sequence, and Eves action. As the analysis will show, we can compare the probability of a particular $y^n$, which is a function of $u^n(m),v^n$ with the number of $y^n$ in $\bigcup_{\hat m \neq m} \typical{Y|U,V}(u^n(\hat m),\emptyset^n) \cap \typical{Y|U,V}(u^n(m),v^n)$ to find as upper bound on the probability of intrusion for a given message. To begin, we first write equation~\eqref{eq:d_1} as a summation of probabilities,
%\begin{align}
%&\sum_{u^n,v^n,y^n } P(y^n,v^n,u^n|M=m) \idc{y^n \in \bigcup_{\hat m \neq m  } \typical{Y|U,\emptyset}( u^n(\hat m))} \label{eq:d_2}.
%\end{align}
Next we seek to determine the probability of error given a particular codebook and $v^n$. This action is equivalent to summing over $y^n$, for equation~\eqref{eq:d_2} can be written
\begin{align}
&\sum_{u^n,v^n,y^n }  P(v^n,u^n|M=m) \notag\\
&\hspace{5pt} \cdot \sum_{y^n} P(y^n|u^n,v^n,M=m) \idc{y^n \!\! \in \!\!\!  \bigcup_{\hat m \neq m  } \!\! \typical{Y|U,\emptyset}( u^n(\hat m))} \label{eq:d_2b}.
\end{align}
Consider $P(y^n|u^n,v^n,M=m)$, for a fixed value $u^n(m),v^n$, there exists an $\epsilon_n \rightarrow 0$ such that 
\begin{equation}\label{eq:ck1}
\Pr \left\{ Y^n \notin \typical{Y|U,V}(u^n(m),v^n) \middle| u^n(m),v^n \right\} \leq \epsilon_n,
\end{equation}
due to~\cite[Lemma~2.12]{CK}.
Furthermore because the distribution $P_{X|U}$ is considered fixed, and because the $\delta_n$ defining the typical set (recall Section~\ref{sec:notation}), there exists a $\lambda_n \rightarrow 0$ such that
\begin{equation}\label{eq:ck2}
\abs{- \frac{1}{n} \logt P(y^n|u^n,v^n,M=m) - H(Y|\tilde U,\tilde V) } \leq \lambda_n 
\end{equation}
where $\tilde U,\tilde V$ are random variables that are jointly distributed according to the type induced from $u^n(m),v^n$, due to~\cite[Problem~2.5.b]{CK}. Because the distribution of $u^n(m),v^n$ is not fixed over all values of $u^n(m)$ and $v^n$, to efficiently apply equations~\eqref{eq:ck1} and~\eqref{eq:ck2} we must partition the values of $(u^n(m),v^n)$ so that over every partition we only need worry about a single distribution. To this end, let the set $\mcf{Q}$ contain all possible joint distributions on $\mcf{U} \times \mcf{V}$. With a slight abuse of notation, a particular realization of $\mcf{Q}$ will be denoted $Q$ in order to emphasize that $Q$ is representative of a certain distribution over $\mcf{U} \times \mcf{V}$. Introducing $Q$ and applying equations~\eqref{eq:ck1} and~\eqref{eq:ck2}, we may upper bound equation~\eqref{eq:d_2b} by
\begin{align}
&\epsilon_n +  \sum_{Q \in \mcf{Q}} \sum_{u^n,v^n : (u^n(m),v^n) \in \type{Q}}  p(u^n,v^n|M=m) \notag \\
&\cdot \sum_{y^n} 2^{-n \left( H(Y|U,V,\mcf{Q}=Q) - \lambda_n \right)} \notag \\
& \cdot \idc{y^n \in  \bigcup_{\hat m \neq m   }  \typical{Y|U,\emptyset}(u^n(\hat m)) \cap \typical{Y|U,V}(u^n(m),v^n)} \label{eq:d_3}.
\end{align}
We have introduced the notation $H(Y|U,V,\mcf{Q} = Q)$ into equation~\eqref{eq:d_3} to make explicit the dependence of the entropy on the distribution of $(u^n(m),v^n)$. Having dealt with the term $P(y^n|u^n,v^n,M=m)$, we turn towards completing the summation. To this end,  
\begin{align}
&\epsilon_n +  \sum_{Q \in \mcf{Q}} \sum_{u^n,v^n : (u^n(m),v^n) \in \type{Q}}  p(u^n,v^n|M=m) \notag \\
&\hspace{3pt} \cdot \sum_{\hat m \neq m} 2^{-n \left( H(Y|U,V,\mcf{Q}=Q) - \lambda_n \right)} \notag \\
& \hspace{3pt}\cdot  \max_{ \begin{matrix} ( U',U,V) : \\ (u^n(\hat m), u^n(m), v^n) \in \type{U',U,V} \\ P_{Y|U'} = P_{Y|U, V = \emptyset} \\ P_{Y|  U, V} = P_{Y|U,V,\mcf{Q} = Q} \end{matrix} } \hspace{-30pt} 2^{ n( H(Y|  U',U,V) + \lambda_n') }  \label{eq:d_4}.
\end{align}
is an upper bound of equation~\eqref{eq:d_3}. Equation~\eqref{eq:d_4} follows first from
\begin{align}
&\idc{y^n \in  \bigcup_{\hat m \neq m}   \typical{Y|U,\emptyset}( u^n(\hat m)) \cap \typical{Y|U,V}(u^n(m),v^n)}\notag \\
&\leq \sum_{\hat m \neq m} \idc{\typical{Y|U,\emptyset}( u^n(\hat m)) \cap \typical{Y|U,V}(u^n(m),v^n)}. \label{eq:d_4b}
\end{align} 
And secondly from~\cite[Problem~2.10]{CK},
\begin{align}
&\sum_{y^n} \idc{y^n \in  \typical{Y|U,\emptyset}(u^n(\hat m)) \cap \typical{Y|U,V}(u^n(m),v^n)} \notag \\
&\leq \max_{ \begin{matrix} ( U',U,V) : \\ (u^n(\hat m), u^n(m), v^n) \in \type{U',U,V} \\ P_{Y|U'} = P_{Y|U, V = \emptyset} \\ P_{Y|  U, V} = P_{Y|U,V,\mcf{Q} = Q} \end{matrix} } \hspace{-30pt} 2^{ n( H(Y|  U',U,V) + \lambda_n') }\label{eq:sumy}.
\end{align}
While it may appear that the bound induced from equation~\eqref{eq:d_4b} introduces a large amount of error, due to channel coding it does not. Recall that $\typical{Y|U,\emptyset}( u^n(\hat m))$ are representative of the region for which the decoder outputs $\hat m$. These regions should rarely overlap, and thus it is most likely that for a $y^n$, there is only a single value of $m$ for which $\idc{\typical{Y|U,\emptyset}( u^n(\hat m)) \cap \typical{Y|U,V}(u^n(m),v^n)}$ is non zero. 

Letting $\mcf{Q}^+$ be the set of all distributions on $\mcf{U} \times \mcf{U} \times \mcf{V}$ which may meet the requirements presented in the max of equation~\eqref{eq:sumy}, and $N_{Q^+}(u^n,v^n,m)$ be the number of $\hat m$ such that $(u^n(\hat m),u^n(m), v^n) \in \type{Q^+}$, we can more succinctly write equation~\eqref{eq:d_4} by
\begin{align}
&\epsilon_n +  \sum_{Q^+ \in \mcf{Q}^+} \sum_{u^n,v^n}  p(u^n,v^n|M=m) N_{Q^+}(u^n,v^n,m) \notag \\
&\hspace{73pt} \cdot  2^{-n \left( I(Y;U'|U,V,\mcf{Q}^+=Q^+) - \tilde \lambda_n \right)}   \label{eq:d_5},
\end{align}
where $\tilde \lambda_n = \lambda_n + \lambda_n'$. The $Q$ term has been removed since it is uniquely determined by $Q^+$. We now turn our attention to giving an upper bound $N_{Q^+}(u^n,v^n,m)$, which will complete the proof. 

In fact, there exists a $\hat \lambda_n \rightarrow 0$ such 
\begin{align}
&\Pr \left\{  N_{Q^+}(U^n,v^n,m) > 2^{n \left( \left| R - I(U';U,V|\mcf{Q}^+ = Q^+) \right|^+ + \hat \lambda_n \right) }  \right\} \notag \\
&\hspace{50pt} \leq e^{-n^2}, \label{eq:hoef}
\end{align}
due to the Hoeffding bound, alternatively~\cite[Lemma~17.9]{CK}. This is because $N_{Q^+}(u^n,v^n,m) = \sum_{\hat m} \idc{ (u^n(\hat m),u^n(m),v^n) \in \type{Q^+}}$, and hence since each codeword is chosen independently from the distribution, $\Pr \left\{  N_{Q^+}(U^n,v^n,m) > \alpha  \right\}$ is simply the probability that the sum of $2^{nR}-1$ independent and identically distributed binomial random variables is greater than $\alpha$. In this case the probability of the random variable can be upper bounded by $2^{-n \left( I(U';U,V| \mcf{Q}^+ = Q^+) - \zeta_n \right)}$ for some $\zeta_n \rightarrow 0$.

Therefore there exists a $\tilde \epsilon_n \rightarrow 0$, $\hat \lambda_n' \rightarrow 0$ such that 
\begin{align}
&\tilde \epsilon_n +  \sum_{Q^+ \in \mcf{Q}^+} \sum_{u^n,v^n}  p(u^n,v^n|M=m)  \notag \\
&\hspace{13pt} \cdot  2^{n \left( \left| R - I(U';U,V|\mcf{Q}^+ = Q^+) \right|^+  - I(Y;U'|U,V,\mcf{Q}^+ = Q^+) + \hat \lambda_n' \right)}   \label{eq:d_6},
\end{align}
upper bounds~\eqref{eq:d_5} since the maximum value of $N_{Q^+}(u^n,v^n,m) < 2^{n R}$, and $2^{nR} e^{-n^2} \rightarrow 0$. By summing over all $u^n,v^n$ and then recognizing since $\mcf{Q}$ is at most polynomial in $n$, we obtain an $\varepsilon_n \rightarrow 0$ such that
\begin{align}
&\tilde \epsilon_n  \notag \\
&+ \max_{Q^+ \in \mcf{Q}^+}   \!\!\! 2^{n \left( \left| R - I(U';U,V|\mcf{Q}^+ = Q^+) \right|^+  - I(Y;U'|U,V,\mcf{Q}^+ = Q^+) + \varepsilon_n \right)}   \label{eq:d_7},
\end{align}
upper bounds equation~\eqref{eq:d_6}.

Equation~\eqref{eq:d_7} leaves us with a bound dependent on the value of $I(U';U,V| \mcf{Q}^+ = Q^+)$ when compared to $R$ for the maximum value of $Q^+ \in \mcf{Q}^+$. Consider the case where $R > I(U';U,V| \mcf{Q}^+ = Q^+)$, where equation~\eqref{eq:d_6} becomes 
\[
\tilde \epsilon_n +  2^{ n \left(  R - I( Y,U,V;U' |\mcf{Q}^+ = Q^+)  + \varepsilon_n \right)}.
\]
For $R < I(Y;U|V = \emptyset) - \varepsilon_n$ this term converges to $0$. Indeed, recall that $P_{Y|U'} = P_{Y|U,\emptyset}$ for all $Q^+ \in \mcf{Q}^+$, and that 
\begin{align*}
I(Y,U,V;U'|\mcf{Q}^+ = Q^+) &\geq I(Y;U'| \mcf{Q}^+ = Q^+) \\
&= I(Y;U|V = \emptyset).
\end{align*}
Conceptually, this bound relates to attacks where Eve does not care what value of $m$ Bob chooses, as long as it is not $\hat m$. 

Now, consider the alternative situation in which $R< I(U';U,V,Y | \mcf{Q}^+ = Q^+)$, for which equation~\eqref{eq:d_7} reduces to
\[
\tilde \epsilon_n + 2^{-n \left( I(Y; U'|U,V,\mcf{Q}^+ = Q^+) - \varepsilon_n \right)},
\] 
which clearly goes to $0$ if $I(Y; U'|U,V,\mcf{Q}^+ = Q^+) \not \rightarrow 0$. But $I(Y;U'|U,V, \mcf{Q}^+ = Q^+) = 0$ if and only if
\begin{align}
 &\sum_{x, u,v} P_{Y|X,V}(y|x,v) P_{X|U}(x|u) P_{U,V|U'}( u,v|u') \notag \\
&\hspace{40pt}= \sum_{x} P_{Y|X,\emptyset}(y|x) P_{X|U}(x|u'), \label{eq:cond}
\end{align}
for the condition implies that $U' \markov U,V \markov Y$. Combining back the two situations, and excluding the distributions of $P_{U,X}$ such that there exists a $V$ satisfying equation~\eqref{eq:cond} gives that on average for our coding scheme $\Pr \left\{ \varphi(Y^n) \neq M \cup \bad \middle|  V^n \neq \emptyset^n \right\} \rightarrow 0$. Which as discussed prior proves the existence of a set of codes which satisfy our inner bound. 

\section{Discussion}

Of critical importance in our result is the ability of Eve to find a distribution $V$ such that $I(Y;U'|U,V) = 0$. By fixing $U=u$, this implies $I(Y;U'|U=u,V) = 0$, which implies the channel must be ``simulatable'' (see~\cite{Binning}) given $U=u$ in order for Eve to manipulate those values of $U$. On the other hand fixing $V=v$ gives that the channel must be manipulable (see~\cite{graves2013coding}) for all values of $V$ Eve chooses to use.

\end{document}